\documentclass[runningheads]{llncs}
\usepackage{graphicx}
\usepackage[ruled,vlined]{algorithm2e}
\usepackage{url}

\usepackage{amsmath}
\DeclareMathOperator*{\argmax}{arg\,max}
\DeclareMathOperator*{\argmin}{arg\,min}

\begin{document}
\title{Towards Dark Jargon Interpretation in Underground Forums}
\author{Dominic Seyler\inst{1} \and 
Wei Liu\inst{1} \and 
XiaoFeng Wang\inst{2} \and 
ChengXiang Zhai\inst{1}}
\authorrunning{D. Seyler et al.}
\institute{University of Illinois at Urbana-Champaign, USA \\ \email{\{dseyler2, weil8, czhai\}@illinois.edu} \and
Indiana University Bloomington, USA \\ \email{xw7@indiana.edu}}
\maketitle              
\begin{abstract}
Dark jargons are benign-looking words that have hidden, sinister meanings and are used by participants of underground forums for illicit behavior. For example, the dark term ``rat'' is often used in lieu of ``\underline{R}emote \underline{A}ccess \underline{T}rojan''. In this work we present a novel method towards automatically identifying and interpreting dark jargons. We formalize the problem as a mapping from dark words to ``clean'' words with no hidden meaning. Our method makes use of interpretable representations of dark and clean words in the form of probability distributions over a shared vocabulary. In our experiments we show our method to be effective in terms of dark jargon identification, as it outperforms another baseline on simulated data. Using manual evaluation, we show that our method is able to detect dark jargons in a real-world underground forum dataset.

\end{abstract}

\section{Introduction}
When bad actors communicate in underground forums (e.g., Silk Road~\cite{silkroad}), they often use jargons to obfuscate their true intentions. They make use of dark jargons, which are benign-looking words that have hidden, sinister meanings, especially among communities in underground forums. For example, when a user posts a thread wanting a ``rat'', what he/she might really want is malware, i.e., ``\underline{R}emote \underline{A}ccess \underline{T}rojan''. As those jargons facilitate an enormous underground economy~\cite{thomas2015framing}, identifying the real meaning of dark words is essential for understanding cybercrime activities and is an important step in order to measure, monitor and mitigate illicit activity.

Recently, there has been substantial research interest in the intersection of Cybersecurity, Information Retrieval~\cite{husari2017ttpdrill,husari2018using,khandpur2017crowdsourcing,liao2016acing,liao2016seeking,mittal2016cybertwitter,mulwad2011extracting,tsai2007detecting}, and Natural Language Processing~\cite{lim2017malwaretextdb,seyler2018identifying,zhou2018automatic,zhu2016featuresmith,zhu2018chainsmith}. However, dark jargon detection and interpretation has not been well studied since only two works are directly related: Yang et al.~\cite{yang2017learn} proposes to detect dark jargon by utilizing a search engine. The authors scrape data from pages that tend to contain dark terms, filter out key words and use the search engine's similar search function to discover new dark words. 
Yuan et al.~\cite{Yuan2018ReadingTC} leverages the context of a word as a representation for the word's meaning. The intuition is that dark words in dark forums appear in drastically different contexts compared to reputable online corpora (e.g., Wikipedia). Dark words are categorized into five general classes. For example, ``blueberry'' is categorized as ``drug'' and not as ``marijuana'', which would be more beneficial for interpretation. We address this limitation as our method provides more interpretable meaning representations by utilizing probability distributions over context words. Another shortcoming of previous approaches is that the actual meaning of the identified dark jargon is mostly unknown. We alleviate this problem by making our framework more expressive and allow dark terms to be mapped to any word/category where the meaning is known. Furthermore, our framework is completely general as it does not require external resources, such as Wikipedia or a search engine.

We formalize the problem of finding underground jargon into a general framework of finding a probabilistic mapping function of dark words to word meanings. We investigate a specific case of this general framework where we find binary mappings of dark words to ``clean'' words, which are words that have no hidden meaning. Further, we develop novel methodology to find dark jargon words in underground forums automatically using the difference in word distributions. This methodology enables us to create interpretable representations of jargon words that can be used to further explain their hidden meanings. In our experiments we make use of a dark corpus of underground forums and evaluate our methodology. We find that our method successfully identifies dark words in a simulated and a real-world setting.

\section{Approach}
\label{sec:dj-approach}

\subsection{General Framework}

In our general framework we use words with no hidden meanings as an direct explanation for the hidden meaning of dark jargon words. Thus, in the most general sense we are interested in a mapping function $hidden\_meaning(V_{dark})$ that takes as input a vocabulary of dark words $V_{dark}$ and outputs a mapping to a vocabulary of ``clean'' words $V_{clean}$, with no hidden meaning. This mapping can be a probability distribution, which expresses the probability of relatedness of a dark word in $V_{dark}$ to all clean words in $V_{clean}$. 

In this work, we investigate the specific case where the probability distribution is forced to have only a single element with probability 1.0. Thus, we are interested in a binary mapping from $V_{dark}$ to $V_{clean}$. However, it is possible to retrieve a more fine-grained distribution, which we leave for future work.

\subsection{Problem Setup}

Our problem setup is as follows: given two text corpora, a dark corpus $C_{dark}$ and a clean corpus $C_{clean}$, the goal is to find the words that are likely to have hidden meanings in the dark corpus and identify their true meaning. We further build a joint vocabulary $V$, which is the most frequent N words from the union of $C_{dark}$ and $C_{clean}$. Then, for each word $w_d \in C_{dark}$ we want to find a word $ w_c \in C_{clean}$, such that $w_c$ expresses the hidden meaning of $w_d$.

We first get a word vector for each word in both corpora, such that every word $w \in V$ has two word vectors $w_{d}$ and $w_{c}$. Second, for each $w_{d}$, we rank all clean word vectors, such that we find the words in $C_{clean}$ that are most similar to $w_{d}$, thereby assuming that the meaning of $w_{d}$ is related to closeness of words in $C_{clean}$ according to some similarity measure. 

We propose to use two methodologies for achieving this mapping. We first introduce a novel method based on word distributions and Kullback-Leibler-divergence~\cite{kullback1951}. We then find another suitable method in cross-context lexical analysis~\cite{massung2017beyond}. In our experiments we compare both methods to understand which one is more performant for our task.

\subsection{Word Distribution Modeling and KL-divergence}
We start by introducing the word distribution and KL-divergence method. The intuition is that a dark word, e.g., ``rat'', will appear in different contexts than the clean word ``rat''. It will therefore have a context more similar to a clean word like ``malware'', as it would have to ``mouse''. When we represent word contexts as probability distributions over words, we find that ``rat'' in the dark corpus and ``malware'' in the clean corpus have the most similar distributions. 

For each word in our vocabulary $V$, we build a unigram probability distribution of all other words in $V$. In order to build this probability distribution we make use of a sliding window technique, where we look at $k$ words before and after the occurence of the word under consideration. We choose to employ this technique, since we are interested in a word's immediate context, as compared to the entire document, which is often used in unigram language modes. 

More specifically, to build a word distribution for a word $w \in V$, we first get a length $|V|$ all zero word count vector, with each entry mapped to a word in $V$. We then go through the whole corpus $C$, and for each occurrence of $w$, we look at $k$ words before and after it, increase the value of the counter vector at corresponding indices. To get a probability distribution over context words, we perform maximum-likelihood estimation and divide each element in the vector by the sum of all vector elements. We further employ smoothing to handle the zero-value probability problem, where we smooth the word distribution of $w$. We get two word distributions for each word $w \in V$:  One distribution estimated from the dark text $P(w_d|C_{dark})$ and one from the clean text $P(w_c | C_{clean})$.
To get two words' dissimilarity $dissim(w_d, w_c)$, we calculate the KL-Divergence between the two probability distributions as in Equation \ref{eq:kl-divergence}. 
Finally, for each dark jargon we define it's hidden meaning as the clean word with the lowest dissimilarity to our target dark word $w_d$ (Equation \ref{eq:min-kl-divergence}).

\begin{equation}
\label{eq:kl-divergence}
    dissim(w_d, w_c) = KL(P(w_d | C_{dark}) || P(w_c | C_{clean}))
\end{equation}

\begin{equation}
\label{eq:min-kl-divergence}
    hidden\_meaning_{KL}(w_d) = \argmin_{w_c \in C_{clean}} dissim(w_d, w_c)
\end{equation}

\subsection{Cross-context Lexical Analysis}

Another suitable method for our problem setup is cross-context lexical analysis~(CCLA)~\cite{massung2017beyond}. Here, the goal is to analyze differences and similarities of words across different contexts. Contexts are usually defined over document collections, which is very akin to our problem setting. Therefore, we can directly apply this methodology to our problem, where the two corpora under consideration are our dark and clean corpora $C_{dark}$ and $C_{clean}$, respectively. Using CCLA as a framework, we can leverage it as yet another method to measure the difference of words in a clean and dark context.

Following Massung~\cite{massung2017beyond}, we define a scoring function as in Equation \ref{eq:ccla-scoring},
where $cos(w_1, w_2, C)$ is the cosine similarity of the word vector of $w_1$ and $w_2$ computed over corpus $C$. $NN(w, C, k)$ is the corresponding length-k vector, where each entry has the value of the cosine similarity of $w$'s word vector and the $k$ closest word vectors. $W_{common}$ is the intersection of the set of $k$ words in corpus C with highest similarity to the word vectors $w_d$ and $w_c$ (Equation \ref{eq:ccla-common-neighbor}).
Note that our function is a slight variation of Massung~\cite{massung2017beyond}, as we modify it to be suitable for two input words ($w_{d}, w_{c}$), rather than just a single input word. Essentially, $\phi$ measures the similarity of the usage of $w_{d}$ and $w_{c}$ across $C_{dark}$ and $C_{clean}$. To generalize, for each word in $w_d \in C_{dark}$, we find a $w_c \in C_{clean}$ that maximizes $\phi$, which is then used as the mapping for $w_d$ (Equation \ref{eq:min-phi}).
\vskip-10pt
\small
\begin{equation}
\label{eq:ccla-scoring}
        \phi(w_{d}, w_{c}, C_{dark}, C_{clean}, k) = \frac{\Sigma_{w \in W_{common}}cos(w, w_d, C_{dark}) * cos(w, w_c, C_{clean})}{||NN(w_{d}, C_{dark}, k)|| * ||NN(w_{c}, C_{clean}, k)||}
\end{equation}
\vskip-10pt

\begin{equation}
\label{eq:ccla-common-neighbor}
        W_{common}(w_{d}, w_{c}, C_{dark}, C_{clean}, k) = W(w_{d}, C_{dark}, k) \bigcap W(w_{c}, C_{clean}, k)
\end{equation}
\vskip-10pt

\begin{equation}
\label{eq:min-phi}
        hidden\_meaning_{CCLA}(w_d) = \argmax_{w_c \in C_{clean}} \phi(w_d, w_c, C_{dark}, C_{clean}, k)
\end{equation}
\normalsize
\vskip-10pt

\section{Experiments}
\label{sec:dj-preliminary-experiments}

\subsection{Experimental Setup}

We aim to answer three research questions: (1) What is the performance of the word distribution method? (2) What is the performance of CCLA compared to the word distribution method? (3) What are the qualitative results in terms of dark jargons identified?

\paragraph{Datasets.} We make use of two datasets in our experiments, where each dataset has stowords and punctuation removed, words are lower-cased and stemmed: (1) \textit{Dark Corpus.} Taken from Yuan et al.~\cite{Yuan2018ReadingTC}, our dark corpus contains user posts scraped from four major underground forums: Silk Road~\cite{silkroad}, Nulled~\cite{nulled}, Hackforums~\cite{hackforums} and Dark0de~\cite{darkode}. The combined corpus contains 376,989 posts.
(2) \textit{Clean Corpus.} The clean corpus contains a web scrape of 1.2 million reddit~\cite{reddit} threads from 1,697 top subreddits in terms the number of subscribers.

\paragraph{Evaluation Environments.} In order to answer our research questions, we build two evaluation environments: The first environment aims to evaluate the quantitative performance of our method. Since no gold standard data is available for this task, we decided to simulate the dark jargons in the dataset. The second environment aims to measure the quality of the dark jargons identified on real data. Here, we manually check if the model can find real meanings of dark words on non-simulated data. The two environments are created as follows:

\noindent (1) \textit{clean-clean}: We randomly split the documents in the clean corpus into two splits. In the first split, namely $\textit{clean}_1$, we randomly select 500 words and prefix them with a dash ("\_"). For example, if the word ``strawberry'' was selected, a sentence like ``John loves \textbf{strawberry} milkshakes'' would be turned into ``John loves \textbf{\_strawberry} milkshakes''. The second split, namely  $\textit{clean}_2$, remains unmodified. Once we run the models on this corpus, for each word in the vocabulary in $\textit{clean}_1$, we get its corresponding ranking list of nearest words in $\textit{clean}_2$. We separately investigate the dashed words (words with "\_"). For those words, the top-ranked word should be the word itself, i.e., the original word without the dash ("\_").  We calculate the mean reciprocal rank (MRR) as a performance evaluation metric for the clean-clean dataset. We separately measure MRR for all words in the vocabulary and for our simulated dark words.

\noindent (2) \textit{dark-clean}: For the real world dataset, we run our word distribution method and get a ranked list of nearest words in $\textit{clean}$ for each word in $\textit{dark}$. We then do a manual evaluation of random dark words our method retrieves to find out their hidden meanings.

\paragraph{Hyperparameters.} We use the following parameters for our methods, which we empirically found to perform best: We use a vocabulary size of 10,000. For the word distribution method, we use a sliding window size $k$ of 10 and $Laplace$\footnote{We found that Dirichlet smoothing was less effective.} smoothing with $\alpha = 1$. For CCLA, we use an embedding size of 300 and a neighborhood size $k$ of 100.

\subsection{Experimental Results}

\begin{table}[t]
	\centering
	\caption{Clean-clean Evaluation of the Word Distribution~(KL) and Cross-context Lexical Analysis~(CCLA) Methods using the Mean Reciprocal Rank (MRR) metric.}
	\label{tab:eval-models}
	\begin{tabular}{c||c | c}
    	\hline
        Method   & MRR all words & MRR dark words  \\
		\hline \hline
		KL  & 0.909 & \textbf{0.892}   \\
		 \hline
		CCLA     & \textbf{0.974} & 0.479  \\
	\end{tabular}
\end{table}

We now move on to our experimental results and answer our three research questions. Table~\ref{tab:eval-models} shows the results of our proposed word distribution method~(KL) and CCLA for all words in the vocabulary and our simulated dark words. To answer our first research question, we see that the KL method performs well, with an MRR around 0.9 for all words in the vocabulary and the simulated dark words. To answer research question two, we find that the CCLA method performs better for all words, however, it is performing much worse for the simulated dark words. Since finding dark words is the goal of our research, we can conclude that KL outperforms CCLA for our task.

To answer research question three, we perform a manual evaluation into the dark words that were identified by our method on a real-world corpus. In Table~\ref{tab:manual-eval}, we present a list of dark words identified by our word distribution method and the clean word that was mapped to the corresponding dark word. We also show the meaning that we manually identified using a slang dictionary or by searching for the highest ranked clean words online. As can be seen from the table, our method retrieves meaningful results since our analysis finds many drug-related and malware-related terms. We take these results as evidence for the potential of our method for finding dark term meanings in a real-world setting.

\begin{table}[t]
	\centering
	\begin{footnotesize}
	\caption{Dark-clean Manual Evaluation based on our Word Distribution Method.}
	\label{tab:manual-eval}
	\begin{tabular}{c | c | c }
    	\hline
        Dark Word & Clean Word  & Meaning \\
		\hline \hline
		gdp & kush &  Grand Daddy Purps (type of marijuana)  \\
		 \hline
		blueberry & kush & type of marijuana  \\
		 \hline
		coke & cocaine & nickname for cocaine  \\
		 \hline
		klonopin & xanax & sedative medication \\
		\hline
		shrooms & lsd & hallucinogenic drug similar to LSD \\
		\hline
		bubba &  kush & type of marijuana \\
		\hline
		ecstasy & mdma  & nickname for mdma \\
		\hline
        dilaudid & oxy, morphine & strong painkiller (aka: hospital heroin) \\
        \hline
        pineapple & kush & type of marijuana \\ 
        \hline
        zeus & botnet & botnet malware \\
        \hline
        rat  & malware & Remote Access Trojan (malware) \\
		\hline
	\end{tabular}
	\end{footnotesize}
\end{table}

\section{Conclusion and Future Work}

We have shown that our approach based on word distributions derived from a word's context is effective for jargon detection and it outperformed a related method based on cross-context lexical analysis. Furthermore, our method leverages word distributions and is therefore inherently interpretable, as individual word probabilities can be thought of as importance weights of a word's context. In the future, we plan to further improve interpretability of dark terms by leveraging external large-scale knowledge resources that define the meaning of slang words, such as Urban Dictionary~\cite{urbandict}.

\section*{Acknowledgments}
This material is based upon work supported by the National Science Foundation under Grant No. 1801652. 
\bibliographystyle{splncs04}
\bibliography{thesisrefs}

\end{document}